\pdfoutput=1

\documentclass[11pt]{article}

\usepackage[final]{acl}

\usepackage{times}
\usepackage{latexsym}
\usepackage{helvet}  
\usepackage{courier}  
\usepackage{natbib}  
\usepackage{caption} 
\usepackage{amsmath}
\usepackage{graphicx}
\usepackage{multirow}
\usepackage{adjustbox}
\usepackage{tabularx}
\usepackage{booktabs}
\usepackage{tablefootnote}
\usepackage{booktabs}
\usepackage{array}
\usepackage{multirow}
\usepackage{color}
\definecolor{lightgray}{RGB}{215,215,215}
\usepackage{colortbl} 
\usepackage{xcolor}
\usepackage{ulem}
\useunder{\uline}{\ul}{}
\usepackage{subfigure}
\usepackage{algorithm}  
\usepackage{algorithmicx}  
\usepackage[noend]{algpseudocode}  

\usepackage{amsmath}  
\usepackage{enumitem}
\usepackage{tabularx}
\usepackage[utf8]{inputenc}
\usepackage[english]{babel}
\usepackage{amsthm}
\usepackage{bm}
\usepackage[T1]{fontenc}

\usepackage[utf8]{inputenc}

\usepackage{microtype}

\usepackage{inconsolata}

\usepackage{graphicx}

\newcommand{\eg}{\textit{e.g., }}

\DeclareMathOperator*{\argmin}{arg\,min}

\title{A Federated Framework for LLM-based Recommendation}

\author{Jujia Zhao$^1$, Wenjie Wang$^{2*}$, Chen Xu$^{3*}$, See-Kiong Ng$^{4}$, Tat-Seng Chua$^{2}$ \\
$^1$Leiden Institute of Advanced Computer Science, Leiden University\\
$^2$NExT++ Research Center, National University of Singapore\\
$^3$Gaoling School of Artificial Intelligence, Renmin University of China\\
$^4$Institute of Data Science and School of Computing, National University of Singapore\\
\texttt{\{zhao.jujia.0913, wenjiewang96\}@gmail.com}\\
\texttt{ \texttt{ xc\_chen@ruc.edu.cn}, \{seekiong, dcscts\}@nus.edu.sg}\\
}

\begin{document}
\maketitle

\let\thefootnote\relax\footnotetext{$^*$Corresponding author.}

\begin{abstract}
Large Language Models (LLMs) have empowered generative recommendation systems through fine-tuning user behavior data.
However, utilizing the user data may pose significant privacy risks, potentially leading to ethical dilemmas and violations of data protection regulations. 
To address the privacy concerns, Federated Learning for Recommendation (Fed4Rec) has been identified as a promising solution. 
However, directly applying Fed4Rec in the LLM context introduces two challenges:
1) exacerbated client performance imbalance, which ultimately impacts the system's long-term effectiveness, and
2) substantial client resource costs, posing a high demand for clients' both computational and storage capability to locally train and infer LLMs.

To tackle these challenges, we propose a federated framework for LLM-based recommendation (shorted as FELLRec).
Generally, FELLRec designs two key strategies. 
1) Dynamic balance strategy, which designs dynamic parameter aggregation and learning speed for different clients, aiming to ensure balanced performance across clients. 
2) Flexible storage strategy, which selectively retains certain sensitive LLM layers on the client side, while offloading other layers to the server, aiming to preserve privacy while saving resources.
Experiment results show that FELLRec can achieve a more balanced client performance and improved overall performance in a computational and storage-efficient way while safeguarding user privacy well.
\end{abstract}
\section{Introduction}
\label{sec:introduction}

Large Language Models (LLMs) with advanced contextual understanding abilities have demonstrated potential in building generative recommendation systems~\cite{rajput2023recommender,gao2023chat}.
Fine-tuning LLMs with user behavior data is essential for learning user preferences~\cite{bao2023bi,li2023text}, 
however, it will face serious privacy leakage risks like in the traditional recommender models. 
The unintended disclosure of sensitive user data could cause ethical issues and infringe upon data protection laws such as the General Data Protection Regulation in the European Union~\cite{hoofnagle2019european}. 
Therefore, ensuring the security and privacy of recommendation data during the LLM fine-tuning process is crucial.

To address the data privacy concerns, Federated Learning for Recommendation (Fed4Rec) emerges as a promising solution~\cite{muhammad2020fedfast,sun2022survey}. 
Fed4Rec requires clients (\eg user devices and platforms with a group of users) to conduct local training using the client's data, and then exchange non-sensitive intermediate parameters such as model parameters and gradients.
This approach protects sensitive user behavior data by keeping them on the client side without the need for sharing with others. 
In General, Fed4Rec mainly employs two frameworks: 
1) Peer-Peer Framework~\cite{yang2022dpmf,an2024nrdl}, which makes every client broadcast the updated parameters to other clients directly within the peer-to-peer network. 
However, this framework faces limitations in LLM-based recommendation scenarios, primarily due to the high communication costs incurred by the large number of LLM parameters.
2) Client-Server Framework~\cite{zhang2023dual,zhang2023lightfr}, which transmits the updated parameters of the clients to a central server for aggregation.  
Previous works~\cite{sun2022survey,yin2024device} have demonstrated that the client-server framework is more efficient in terms of communication overhead, making it ideal for LLM-based recommendations.

\begin{figure}
\setlength{\abovecaptionskip}{-0.10cm}
\setlength{\belowcaptionskip}{-0.4cm}
  \centering 
  \hspace{-0.2in}
  \subfigure{
    \includegraphics[width=1.5in]{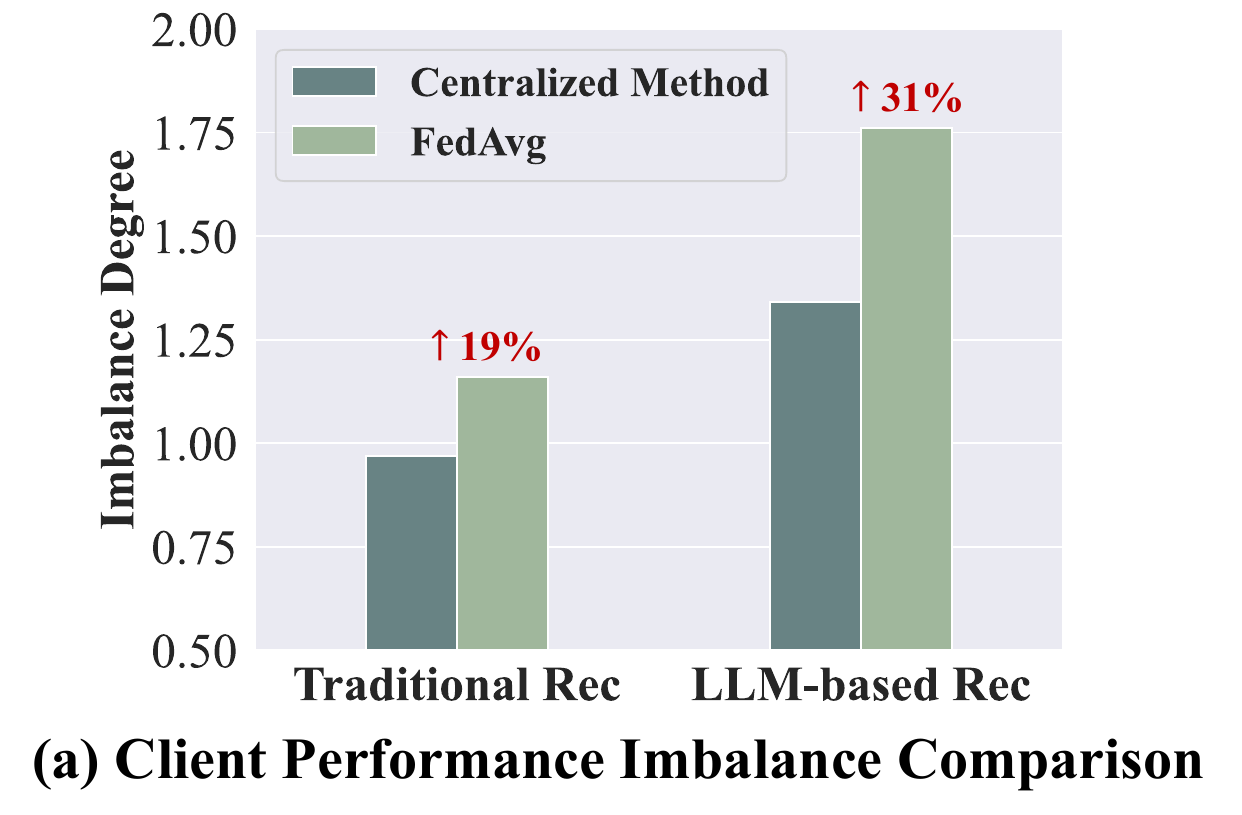}} 
  \hspace{-0.1in}
  \subfigure{
     \includegraphics[width=1.5in]
     {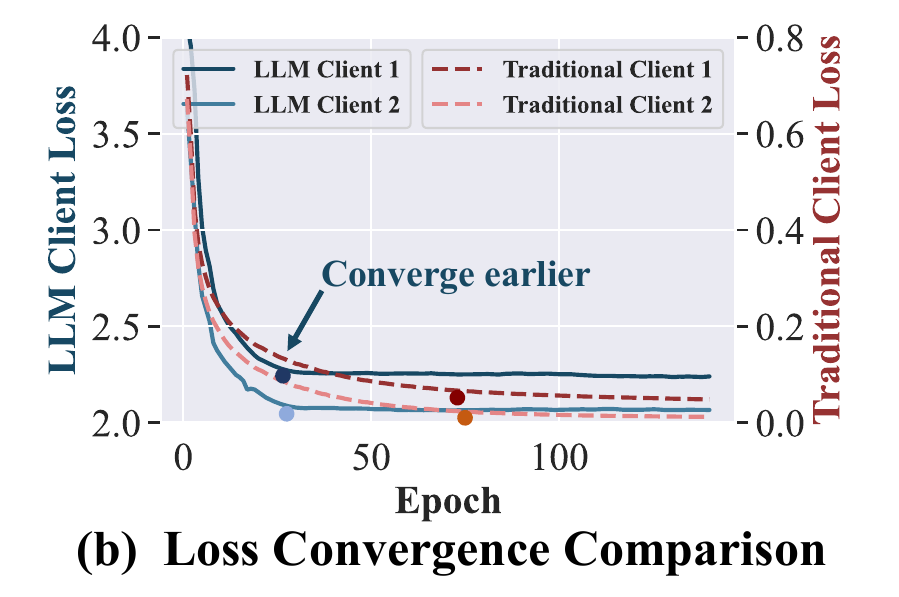}} 
\caption{(a) illustrates the exacerbated client performance imbalance when applying a classical client-server method (FedAvg~\cite{mcmahan2017communication}) to LLM-based recommender models (BIGRec) compared with traditional recommender models (MF). (b) shows the convergence rate of two selected clients when applying FedAvg to LLM-based and traditional models. The observations are on Games.}
  \label{fig:intro}
\end{figure}

However, adapting the client-server framework to LLM-based recommendation presents two challenges: 
1) \textbf{Exacerbated Client Performance Imbalance}: 
Based on our empirical analysis in Figure~\ref{fig:intro}(a), it is evident that directly applying the client-server framework to LLM-based recommendation models leads to a more significant client performance imbalance compared to traditional models. 
This exacerbated imbalance may cause less accurate and equitable recommendations for specific clients, ultimately impacting the system's long-term effectiveness and user satisfaction~\cite{xu23-pmmf, burke2018balanced}. 
This exacerbated imbalance potentially stems from the accelerated training convergence among clients, as depicted in Figure~\ref{fig:intro}(b), possibly due to the fast adaptation capabilities of LLMs~\cite{bao2023bi,bao2023tallrec}. 
2) \textbf{Substantial Client Resource Costs}:
The client-server framework necessitates that each client possesses the capability to locally train and infer LLMs. However, the extensive computational and storage resources required by LLMs pose a substantial challenge for individual clients in meeting these demands~\cite{chen2023federated,fan2023fate}.

To tackle the issues of exacerbated performance imbalance and substantial resource costs, we refine the client-server framework with two strategies: 
1) \textbf{Dynamic Balance Strategy}. 
To mitigate the performance imbalance among clients, we introduce a dynamic balance strategy:
it involves designing dynamic parameter aggregation and learning speed for each client to ensure relatively equitable performance across the board.
2) \textbf{Flexible Storage Strategy}: 
To reduce client costs, we propose a flexible storage strategy for the client model. 
Intuitively, this strategy selectively allocates some LLM layers, especially those capable of extracting sensitive user data, on the client side, while situating other non-sensitive layers on the server to save cost. 

In light of these, we propose a \textbf{Fe}derated Framework for \textbf{LL}M-based Recommendation (FELLRec). 
1) FELLRec adapts dynamic balance strategies for different clients. 
Specifically, FELLRec preserves personalized parameters on each client (\eg Low-Rank Adaption (LoRA)~\cite{hu2021lora}) and employs a dynamic parameter aggregation method based on attention mechanisms. 
Meanwhile, FELLRec devises dynamic learning speed by proposing a Curriculum Heating learning method~\cite{chen2021curriculum} based on client loss, which helps client undergoes a gradual pre-warming phase to familiarize themselves with their own data distribution.
2) FELLRec adopts the flexible storage strategy to deploy those input and output layers on the client side to ensure the protection of all sensitive information (see detailed analysis in Section~\ref{sec:tradeoff}).
Empowered with the two strategies, FELLRec can safeguard data privacy for LLM-based recommendations in a more balanced and efficient way. We instantiate FELLRec on two LLM backend models and conduct extensive experiments on three datasets, validating its effectiveness and efficiency. 

\textbf{Main Contributions:}
(1) We introduce a privacy-preserving task for fine-tuning LLM-based recommendation models, where we identify the challenges of directly adopting Fed4Rec: exacerbated client performance imbalance and substantial client resource costs.  
(2) We propose a federated framework for LLM-based recommendation called FELLRec, which addresses the two challenges well while preserving data privacy. 
(3) Experiments across three public datasets under various settings, confirming its efficacy and efficiency\footnote{Our code and data are released at \url{https://github.com/Polaris-JZ/FELLRec}.}. 

\section{Preliminary }
\label{sec:preliminary}

\subsection{LLM-based Recommendation}
Let $\mathcal{U}$, $\mathcal{I}$ be the user set and item set, 
for a given user $u\in\mathcal{U}$, the LLM-based recommender $f(\bm{\mathcal{P}})$ will utilize 
the user's historical interactions $H_u$ 
to generate a ranking list $L_K(H_u)\subset\mathcal{I}$ as the recommendation for user $u$, where $K$ is the item numbers in a ranking list and $\bm{\mathcal{P}}$ is the parameter set of LLMs. 
$H_u$ is the user $u$'s browsing history: $H_u = [i_1, i_2, \cdots, i_N]$, where $i_n\in\mathcal{I}$ is the $n-$th item in the interaction history (typically in a natural language form), and $N$ is the history length.


\subsection{Client-Server Framework under Fed4Rec}

Let $\mathcal{C}$ be the client set, where each client $c\in\mathcal{C}$ could be a user $u$ or a group of users from a specific platform. Each client $c$, equipped with a model parameter $\bm{\mathcal{P}}_c$, has a local dataset $\bm{\mathcal{D}}_c = \{(H_u, y), \forall u\in c\}$, which includes the users' interaction history $H_u$ and the label $y$ (usually the next-interacted item) for training. 
Within the client-server framework, the most classic approach is FedAvg~\cite{mcmahan2017communication}.
Specifically, at each training epoch, FedAvg utilizes a central server to aggregate parameters from various clients to generate unified updated parameters for every client. 
Formally, at each epoch, each client is required to update its parameter based on its local dataset:
$
    \bm{\mathcal{P}}_c = \argmin_{\bm{\mathcal{P}}_c} \sum_{(H_u,y)\in\bm{\mathcal{D}}_c}l(f(H_u; \bm{\mathcal{P}}_c), y), \forall c \in\mathcal{C},
$
where $l(\cdot)$ is the loss function of recommendation. 
Subsequently, the central server will aggregate parameters from all clients, and send the unified aggregated parameters back to every client:
$
    \bm{\mathcal{P}}_c = \dfrac{1}{n} \sum_{c'\in \mathcal{C}}\frac{|\bm{\mathcal{D}}_{c'}|}{|\bm{\mathcal{D}}|} \mathcal{P}_{c'}, \forall c\in \mathcal{C}.
$
FedAvg ensures privacy by obviating the necessity to transmit original data to the server, concentrating instead on the exchange of model parameters. 
However, directly applying FedAvg to LLM-based recommendations will meet the exacerbated client performance imbalance and substantial client resource cost. 
\section{FELLRec}
\label{sec:method}

\begin{figure*}
\setlength{\abovecaptionskip}{0.03cm}
\setlength{\belowcaptionskip}{-0.2cm}
\centering
\includegraphics[scale=0.65]
{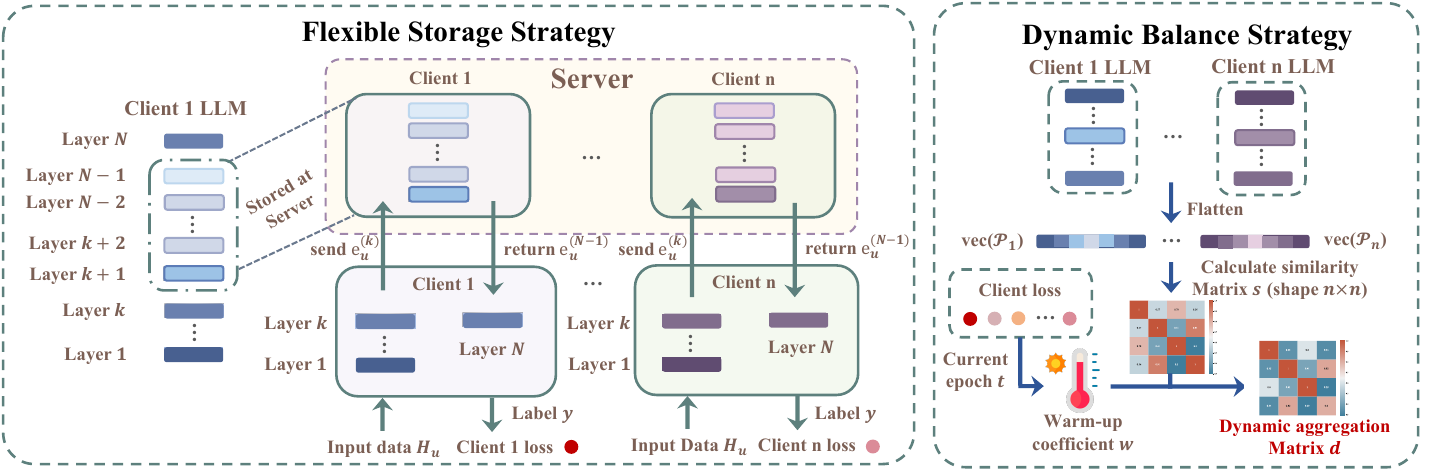}
\caption{FELLRec Structure. The left part is the flexible storage strategy which offloads non-sensitive LLM layers to the server to save resources. The right part is the dynamic balance strategy which ensures relatively balanced performance across clients.}
\label{fig:method}
\end{figure*}

In response to exacerbated client performance imbalance and significant client resource costs, we introduce a Federated Framework for LLM-based Recommendation (FELLRec), which enhances data privacy in LLM-based recommendation systems both equitably and efficiently. 
FELLRec encompasses two strategies:
1) Dynamic balance strategy, which designs dynamic parameter aggregation and learning speeds to ensure relatively balanced performance across clients.
2) Flexible storage strategy, which enables the flexible storage of LLM layers to conserve resources. 
The architecture of FELLRec is depicted in Figure~\ref{fig:method}.

\subsection{\textbf{Dynamic Balance Strategy}}\label{sec:balance}
As illustrated in Section~\ref{sec:introduction}, directly applying the client-server framework in LLM-based recommendation exacerbates the performance imbalance across clients.
This imbalance may lead to less equitable recommendations for specific clients, thereby detrimentally affecting the system's overall effectiveness and diminishing user satisfaction.

The imbalance could be potentially attributed to two primary factors:
1) the diverse data distribution among clients, which may lead to conflicting optimization objectives among clients, thus possibly sacrificing the performance of specific clients.
2) The varied learning difficulty levels among clients, where those facing greater challenges may exhibit relatively poorer performance.

To address these issues, FELLRec first ensures client personalization by maintaining client-specific parameters for each client, including two kinds:
1) LoRA~\cite{hu2021lora}, and
2) either a part or the entirety of LLM's own parameters.
To economize on client resources, the remaining parameters are fixed. 
Our analysis primarily utilizes LoRA as an illustrative example, as the same principles apply to other methods.

Specifically, for client $c$, the model parameters are denoted as $\bm{\mathcal{P}}_c$ with LoRA $\bm{\mathcal{R}}_c$, where LoRA is client-specific parameters and $\bm{\mathcal{P}}_c$ is the fixed original LLM model parameters.
Subsequently, FELLRec incorporates a dynamic balance strategy, which involves designing dynamic parameter aggregation and learning speeds for each client, addressing two key factors of imbalance respectively.

An intuitive idea of our method is:
\begin{equation}
\label{eqn:m1}
\scalebox{0.8}{  
$\bm{\mathcal{R}}_c = \dfrac{\sum_{c'\in \mathcal{C}} d_{c,c'} \bm{\mathcal{R}}_{c'}}{\sum_{c'\in \mathcal{C}} d_{c,c'}}$
}
\end{equation}
where $d_{c,c'}$ is the dynamic aggregation weight and it can be divided into two parts: $d_{c,c'}=w_c\cdot s_{c,c'}$, where $ s_{c,c'}$ is the attention-based aggregation weight corresponding to the Sub-section~\ref{sec:aggregation} and $w_c$ is the learning difficulty weight illustrated at the Sub-section~\ref{sec:speed}. 

\subsubsection{\textbf{Dynamic Parameter Aggregation}}\label{sec:aggregation}
Given the variability in data distribution among clients, the optimization objectives for them may diverge, potentially leading to conflicts when trying to optimize a global model.
Such conflicts may inadvertently sacrifice the performance of specific clients, which causes imbalance.

Given this, FELLRec incorporates an attention-based parameter aggregation method.
This method customizes the aggregation process of each client according to their unique data distribution, aiming to mitigate performance imbalances without compromising the performance of specific clients. 
Intuitively,
the client model prioritizes learning from clients with similar data distributions while reducing the influence of those deemed non-relevant. 
The prioritization mechanism involves aggregating the client model parameters based on the cosine similarity between the parameters of the current client and those of other clients. 
Specifically, for client $c$, the aggregation formula is:
\begin{equation}
\label{eqn:m2}
\scalebox{0.8}{
$ \bm{\mathcal{R}}_c = \dfrac{\sum_{c'\in \mathcal{C}} s_{c,c'} \bm{\mathcal{R}}_{c'}}{\sum_{c'\in \mathcal{C}} s_{c,c'}}$}, \quad \rm where\\ 
\end{equation}
\begin{equation}
\label{eqn:m3}
\scalebox{0.8}{
$ s_{c,c'} = \dfrac{\text{vec}(\bm{\mathcal{R}}_c)^{\top} \text{vec}(\bm{\mathcal{R}}_{c'})}{\|\text{vec}(\bm{\mathcal{R}}_c)\|_2 \|\text{vec}(\bm{\mathcal{R}}_{c'})\|_2}$}. \\ 
\end{equation}
$\|\cdot\|_2$ denotes the $\ell_2$ norm, $\bm{\mathcal{R}}_c$ represents the client-specific LoRA parameter of client $c$, $\text{vec}(\bm{\mathcal{R}}_c)$ represents flattened one-dimension client-specific LoRA parameter of client $c$, and $s_{c,c'}$ is attention-based aggregation weight, which is cosine similarity between $\text{vec}(\bm{\mathcal{R}}_c)$ and $\text{vec}(\bm{\mathcal{R}}_{c'})$.

Through dynamic parameter aggregation, FELLRec ensures more balanced performance across clients by customizing the aggregation process of each client based on its specific data distribution.

\subsubsection{\textbf{Dynamic Learning Speed}}\label{sec:speed}
Given the varied heterogeneity within client datasets, clients encounter different learning difficulties during training~\cite{yang2023fedrich}. 
Consequently, the learning status of different clients (\eg ongoing learning, convergence or overfitting) can vary significantly. 
If a client has not adequately learned from its own data, excessively aggregating parameters from other clients may detrimentally affect its performance, potentially leading to performance imbalances across clients.

In response to this challenge, we develop a client-specific dynamic learning speed mechanism. 
This mechanism dynamically adjusts the extent of learning from other clients according to the client's current learning status, thereby personalizing the client's learning process.
FELLRec assesses a client's learning status via its local loss, which serves as a gauge of the client's learning difficulty, and adjusts the extent of learning from peers accordingly. 
Based on this, FELLRec introduces a Curriculum Heating learning method~\cite{chen2021curriculum}, which is adapted based on client loss.
Intuitively, clients experiencing higher losses undergo a gradual pre-warming phase, allowing them to acclimate to their data distribution, whereas clients with lower losses engage in a rapid convergence, enhancing training efficiency.
Specifically, for client $c$, warm-up coefficient is:
\begin{equation}
\label{eq:m4}
\scalebox{0.8}{
$ w_{c} = \text{tanh}(\dfrac{\alpha}{\left[\exp(\mathcal{L}_c)/\sum_{i=1}^N \exp(\mathcal{L}_i))\right]^{t/\beta}})$}, \\
\end{equation}

where $\alpha$ is the speed-related warm-up factor, influencing the warm-up's overall pace; $\beta$ 
is the time-related warm-up factor, affecting the 
temporal impact on warm-up speed.
In essence, a higher $\alpha$ or a lower $\beta$ accelerates warm-up for clients. 
$w_{c}$ is posed on the similarity score with other clients to control the learning speed:
$
d_{c,c'} =  w_{c} s_{c,c'}, \quad \forall c'\in\mathcal{C}, c'\neq c,
$
where $d_{c,c'}$ is the final dynamic aggregation weight.
This approach, through the application of the warm-up coefficient, dynamically adjusts a client's learning pace based on its current learning status, providing a tailored learning speed for each client and mitigating performance imbalances across the clients.

\subsection{Flexible Storage Strategy}\label{sec:storage} 
In LLM-based recommendation systems, training and inference processes demand significant resource investment. 
Recognizing that not all clients have the capacity for the storage and computational demands of an LLM model, FELLRec introduces a flexible storage strategy aimed at reducing resource expenditure for clients. 

FELLRec retains specific subsets of layers on the client side\footnote{The practical application of Apple Inc. demonstrates the feasibility of deploying LLMs, such as those with 3 billion or 7 billion parameters, on the client side.~\cite{2024Apple}}, particularly those closer to the input and output layers, due to their processing of sensitive data. The rest of the layers are hosted on the server side to save resources.
Speicfically, for client $c$, the client model parameters are denoted as $\bm{\mathcal{P}}_c$ with LoRA $\bm{\mathcal{R}}_c$.
Both of this two parts $\bm{\mathcal{P}}_c$ and $\bm{\mathcal{R}}_c$ can be divided into $N$ layers: $\{\bm{\mathcal{P}}_c^{(i)}\}_{i=1}^N$, $\{\bm{\mathcal{R}}_c^{(i)}\}_{i=1}^N$, respectively, where $N$ is the total number of LLM layers.
Therefore, we combine them as $\bm{\mathcal{T}}_c$ for simplicity, where $\bm{\mathcal{T}}_c = \{\bm{\mathcal{P}}_c^{(i)}, \bm{\mathcal{R}}_c^{(i)}\}_{i=1}^N$
Based on this,
The layer retained on the client side are $\{\bm{\mathcal{T}}_c^{(i)}\}_{i=1}^k$ and $\bm{\mathcal{T}}_c^{(N)}$, where $k$ represents the layer-allocation hyper-parameter.
Conversely, the layers stored on the server side are $\{\bm{\mathcal{T}}_c^{(i)}\}_{i={k+1}}^{N-1}$.

During each training round, the client sends input data $H_u$ to its preserved input layers, which then forwards the output embedding $\bm{e}_u^{(k)} = g(H_u, \{\bm{\mathcal{T}}_c^{(i)}\}_{i=1}^k)$ 
to the server for further processing, 
where $g(\cdot)$ commonly is the attention layer with feed-forward layer in the LLM scenario. 
The server processes this embedding and returns the output $\bm{e}_u^{(N-1)} = g(\bm{e}_u^{(k)}, \{\bm{\mathcal{T}}_c^{(i)}\}_{i={k+1}}^{N-1})$ to the client to produce the final output embedding $\bm{e}_u^{N} = g(\bm{e}_u^{(N-1)}, \bm{\mathcal{T}}_c^{(N)})$. 
Subsequently, the client calculates the loss using the preserved label on the client side.
Formally, the forward process of FELLRec is described as $\bm{\mathcal{T}_c} = \{\bm{\mathcal{T}}_c^{(i)}\}_{i=1}^k \circ \{\bm{\mathcal{T}}_c^{(i)}\}_{i={k+1}}^{N-1} \circ \bm{\mathcal{T}}_c^{(N)}$, where $\circ$ represents operation composition, with the output of the function on the right being used as the input to the function on the left.
Following this, the backward process begins, with gradients propagated in reverse: from the client to the server and then back to the client.


When client and server configurations are consistent, FELLRec's performance is unaffected by the parameter $k$ since offloading layers to the server doesn't alter the training mechanics—only the storage location of model parameters changes. Specifically, both the forward and backward propagation processes proceed identically to scenarios where no layers are offloaded to the server.

This strategy significantly reduces client resource costs during both training and inference, as shown in Table~\ref{tab:cost}. 
It is noteworthy that the determination of the number of layers to preserve is adaptable, enabling control over client costs.
However, despite our method's efforts to protect data privacy, there may be malicious behavior from the server side aimed at attacking the model to access user privacy data. 
Our subsequent experiments indicate that retaining more layers on the server side increases the vulnerability to attacks (as detailed in Section~\ref{sec:tradeoff}), where we analyze the trade-off between the risk of attacks and the costs.

\begin{algorithm}[t]
	\caption{\textbf{FELLRec Training Phase}}  
	\label{algo:train}
	\begin{algorithmic}[1]
		\Require The client set $\mathcal{C}$, item set $\mathcal{I}$, epoch number $T$, local round number $R$, warm-up parameter $\alpha, \beta$, personalized parameters $\mathcal{R}_c$ and local data $\mathcal{D}_c = \{H_u, y\}$, $\forall c\in\mathcal{C}$.
        \Ensure Fine-tuned personalized parameters $\mathcal{R}_c, \forall c\in\mathcal{C}$.
        \State Initialize client model $\mathcal{R}_c, c\in\mathcal{C}$
        \ForAll{each epoch $t = 1, 2, \cdots, T$}
        \State Initialize $\mathcal{L}_c = 0$ 
        \State $// ~~\texttt{Client Local Training}$
            \ForAll{each client $c \in \bm{\mathcal{C}}$ in parallel}
                \ForAll{each round $r = 1, 2, \cdots, R$}
                \State \resizebox{0.7\hsize}{!}{$l_c(\bm{\mathcal{R}_c})=\sum_{(H_u,y)\in\mathcal{D}_c}l(f(H_u; \mathcal{R}_c), y)$}
                \State \resizebox{0.25\hsize}{!}{$\mathcal{L}_c = \mathcal{L}_c + l_c$}
                \State \resizebox{0.39\hsize}{!}{$\mathcal{R}_c = \argmin_{\bm{\mathcal{R}}_c} l_c(\mathcal{R}_c)$}
                \EndFor
            \EndFor
        \State Upload $\{\bm{\mathcal{R}_c^{(i)}}\}_{i=1}^{k}, \bm{\mathcal{R}_c^{(N)}}, \forall c$ to server 
        \State $// ~~\texttt{Aggregate parameters for clients}$
        \State 
         \resizebox{0.7\hsize}{!}{$w_{c} = \text{tanh}(\dfrac{\alpha}{\left[\exp(\mathcal{L}_c)/\sum_{i=1}^N \exp(\mathcal{L}_i))\right]^{t/\beta}})$}
         \State \resizebox{0.75\hsize}{!}{$s_{c,c'} = \dfrac{\text{vec}(\bm{\mathcal{R}}_c)^{\top} \text{vec}(\bm{\mathcal{R}}_{c'})}{\|\text{vec}(\bm{\mathcal{R}}_c)\|_2 \|\text{vec}(\bm{\mathcal{R}}_{c'}\|_2)}, \forall c, c'\in\mathcal{C}$}
         \State \resizebox{0.25\hsize}{!}{$d_{c,c'}=w_{c}s_{c,c'}$}
         \State \resizebox{0.47\hsize}{!}{$\bm{\mathcal{R}}_c = \dfrac{\sum_{c'\in \mathcal{C}} d_{c,c'} \bm{\mathcal{R}}_{c'}}{\sum_{c'\in \mathcal{C}} d_{c,c'}}, \forall c\in\mathcal{C}$}
        \State Send \resizebox{0.43\hsize}{!}{$\{\bm{\mathcal{R}_c^{(i)}}\}_{i=1}^{k}, \bm{\mathcal{R}_c^{(N)}}, \forall c\in\mathcal{C}$} back to clients
        \EndFor
    \end{algorithmic}
\end{algorithm}
\setlength{\textfloatsep}{0.28cm}

\begin{algorithm}[t]
	\caption{\textbf{FELLRec Inference Phase}}  
	\label{algo:infer}
	\begin{algorithmic}[1]
		\Require The client set $\mathcal{C}$, item set $\mathcal{I}$, ranking size $K$, parameters of each client $\bm{\mathcal{T}}_c, \forall c\in\mathcal{C}$, the user $u$
        \Ensure Ranking list $L_K(u)$ 
        \State $// ~~\texttt{Offline Storage}$
        \ForAll{client $c \in \mathcal{C}$}
            \State Get item embeddings \resizebox{0.41\hsize}{!}{$\bm{e}_c(i) = f(i,\{\bm{\mathcal{T}}_c^{(i)}\}_{i=1}^N), \forall i$}
        \EndFor
        \State User $u$ arrives in FELLRec;
        \State Finding $u$ corresponds to the client $c$;
        \State $// ~~\texttt{Client c executes}$
        \State \resizebox{0.45\hsize}{!}{$\bm{e}_u^{(k)}=g(H_u, \{\bm{\mathcal{T}}_c^{(i)}\}_{i=1}^k)$}
        \State Upload $e_u^{(k)}$ to Server;
        \State $// ~~\texttt{Server executes}$
        \State \resizebox{0.57\hsize}{!}{$\bm{e}_u^{(N-1)}=g(\bm{e}_u^{(k)}, \{\bm{\mathcal{T}}_c^{(i)}\}_{i=k+1}^{N-1})$}
        \State Upload $\bm{e}_u^{(N-1)}$ to Client;
        \State $// ~~\texttt{Client c executes}$
        \State Get output embedding \resizebox{0.45\hsize}{!}{$\bm{e}_u^{(N)}=g(\bm{e}_u^{(N-1)}, \{\bm{\mathcal{T}}_c^{(N)})\}$}
        \State $// ~~\texttt{Ranking Step}$
        \State \resizebox{0.9\hsize}{!}{$L_K(u) = \argmin_{S \subset \mathcal{I}, |S| = K} \sum_{i\in S} \text{distance}(\bm{e}_c(i),\bm{e}_u^{(N)})$}
    \end{algorithmic}
\end{algorithm}
\setlength{\textfloatsep}{0.28cm}

\subsection{FELLRec Framework}\label{sec:framework} 

\subsubsection{\textbf{Training}}

In the training phase, FELLRec trains personalized parameter $\bm{\mathcal{R}}_c$ for each client $c$ without sharing their data. Specifically, at each epoch $t$, FELLRec first conducts client local training and then aggregates parameters of all clients to update their personalized parameter $\bm{\mathcal{R}}_c$.

Specifically, at the client local training phase, each client updates their parameters $\bm{\mathcal{R}}_c$ utilizing their respective local datasets $\bm{\mathcal{D}}_c$. 
During this phase, the client model is not entirely stored on the client side. Instead, parts of the model are stored on the server side, as dictated by the flexible storage strategy, to reduce the resource costs associated with training LLMs (see detailed analysis Section~\ref{sec:storage}).
Subsequently, each client uploads their client-preserved parameters to the server for aggregation, making use of the parameters from other clients to assist the update process.

In the aggregate phase, each client gets their dynamic aggregation weight $d_{c,c'}, \forall c'\in\mathcal{C}$ through dynamic parameter aggregation and dynamic learning speed. 
Subsequently, they get aggregated personalized parameters $\bm{\mathcal{R}}_c$ based on their specific aggregation weight and then send client-preserved parameters back to clients. 
The training algorithm of FELLRec is provided in Algorithm~\ref{algo:train}.

\subsubsection{\textbf{Inference}}

    In the inference phase, for any given client $c$, FELLRec utilizes the updated LoRA parameters $\bm{\mathcal{R}}_c$ and fixed parameters $\bm{\mathcal{P}}_c$ to form the complete parameters $\bm{\mathcal{T}}_c$, and then get ranking list $L_K$ as recommendation for user $u$ belongs to this client. 
    
    Specifically, the inference phase is divided into four phases: 
    1) Client $c$ independently stores the embeddings $\bm{e}_c(i)$ for all items from the item corpus $\mathcal{I}$, in preparation for the ranking step. 
    2) Client $c$ gets the hidden embedding at $k-$th layer of LLM through: $\bm{e}_u^{(k)}=g(H_u, \{\bm{\mathcal{T}}_c^{(i)}\}_{i=1}^k)$; 
    3) Then the server receives the uploaded $\bm{e}_u^{(k)}$ and continue to compute the hidden embedding $\bm{e}^{(N-1)}$ at $(N-1)-$th layer of LLM through $\bm{e}_u^{(N-1)}=g(\bm{e}_u^{(k)}, \{\bm{\mathcal{T}}_c^{(i)}\}_{i=k+1}^{N-1})$; 
    4) Finally, client $c$ directly computes the distance (\eg cosine similarity\cite{bao2023bi} or L2 distance~\cite{li2023text}) between the generated embedding $\bm{e}_u^{(N)}$ and the item embeddings $\bm{e}_c(i)$ from item corpus, and get the final ranking list through:
    $
        L_K(u) = \argmin_{S \subset \mathcal{I}, |S| = K} \sum_{i\in S} \text{distance}(\bm{e}_c(i),\bm{e}_u^{(N)}).
    $

    This approach preserves sensitive user data on the client side during inference, thus enhancing data privacy. 
    Additionally, by offloading portions of the computation to the server, FELLRec reduces the computational load on clients and minimizes their hardware requirements.
    The inference algorithm of FELLRec is provided in Algorithm~\ref{algo:infer}.

\section{Experiments}
\label{sec:experiment}
In this section, we conduct a comprehensive experimental study to analyze the performance of FELLRec and the impact of different components (\eg dynamic balance strategy and flexible storage strategy) within it.

\begin{table*}[t]
\setlength{\abovecaptionskip}{0cm}
\setlength{\belowcaptionskip}{0cm}
\caption{Overall performance of FELLRec and other baselines in the LLM-based Recommendation scenario. Bold signifies the best performance among the privacy-preserving methods under the same backend models. * denotes statistically significant improvements of FELLRec over the second-best privacy-preserving methods under the same backend models, according to the t-tests with a significance level of $p$ \textless 0.01.}
\begin{adjustbox}{width={\linewidth},keepaspectratio}
\label{tab:main_exp}
\begin{tabular}{l|cccc|cccc|cccc}
\toprule
                                  & \multicolumn{4}{c|}{\textbf{Games}}                                  & \multicolumn{4}{c|}{\textbf{Microlens}}                           & \multicolumn{4}{c}{\textbf{Book}}                             \\
\multirow{-2}{*}{\textbf{Method}} & \textbf{R@10}   & \textbf{R@20}   & \textbf{N@10}   & \textbf{N@20}   & \textbf{R@10}   & \textbf{R@20}   & \textbf{N@10}   & \textbf{N@20}   & \textbf{R@10}   & \textbf{R@20}   & \textbf{N@10}   & \textbf{N@20}   \\ \midrule
\textbf{BIGRec}                    & 0.0194               & 0.0316               & 0.0127               & 0.0164          & 0.0089               & 0.0132               & 0.0050                & 0.0062      & 0.0079               & 0.0097               & 0.0126               & 0.0116               \\
\textbf{$\quad$+FedAvg}              & 0.0145               & 0.0257               & 0.0093               & 0.0126        & 0.0021               & 0.0039               & 0.0012                & 0.0017                & 0.0081               & 0.0097               & 0.0119               & 0.0112               \\
\textbf{$\quad$+FedProx}     & 0.0143 & 0.0255 & 0.0090 & 0.0123 & 0.0033 & 0.0051 & 0.0032 & 0.0040 & 0.0081 & 0.0096 & 0.0120 & 0.0112 \\
\textbf{$\quad$+Ditto}     & 0.0147 & 0.0260 & 0.0091 & 0.0126 & 0.0040 & 0.0063 & 0.0041 & 0.0045 & 0.0077 & 0.0091 & 0.0113 & 0.0106 \\
\textbf{$\quad$+RoLoRA}     & 0.0128 & 0.0231 & 0.0079 & 0.0106 & 0.0019 & 0.0037 & 0.0013 & 0.0019 & 0.0052 & 0.0075 & 0.0101 & 0.0098 \\
\textbf{$\quad$+Ours}             & \textbf{0.0158*}      & \textbf{0.0274*}      & \textbf{0.0104*}      & \textbf{0.0139*}     & \textbf{0.0088*}      & \textbf{0.0128*}      & \textbf{0.0051*}      & \textbf{0.0062*}    & \textbf{0.0085*}      & \textbf{0.0102*}      & \textbf{0.0124*}      & \textbf{0.0116*}      \\ \midrule
\textbf{RecFormer}                 & 0.0193               & 0.0360                & 0.0117               & 0.0169       & 0.0190                & 0.0369               & 0.0104               & 0.0155            & 0.0318               & 0.0512               & 0.0333               & 0.0380                \\
\textbf{$\quad$+FedAvg}             & 0.0149               & 0.0262               & 0.0089               & 0.0124      & 0.0086               & 0.0192               & 0.0045               & 0.0074                    & 0.0078               & 0.0123                & 0.0085               & 0.0097               \\
\textbf{$\quad$+FedProx}     & 0.0150 & 0.0266 & 0.0086 & 0.0121 & 0.0086 & 0.0166 & 0.0041 & 0.0064 & 0.0071 & 0.0061 & 0.0083 & 0.0133 \\
\textbf{$\quad$+Ditto}     & 0.0162 & 0.0273 & 0.0091 & 0.0138 & 0.0091 & 0.0172 & 0.0046 & 0.0065 & 0.0102 & 0.0131 & 0.0107 & 0.0159 \\
\textbf{$\quad$+RoLoRA}     & 0.0132 & 0.0257 & 0.0081 & 0.0118 & 0.0084 & 0.0187 & 0.0029 & 0.0045 & 0.0071 & 0.0115 & 0.0079 & 0.0095 \\
\textbf{$\quad$+Ours}           & \textbf{0.0215*}      & \textbf{0.0373*}      & \textbf{0.0122*}      & \textbf{0.0170*}    & \textbf{0.0141*}      & \textbf{0.0245*}      & \textbf{0.0065*}      & \textbf{0.0094*}        & \textbf{0.0274*}      & \textbf{0.0411*}      & \textbf{0.0275*}      & \textbf{0.0301*} \\ \bottomrule
\end{tabular}
\end{adjustbox}
\end{table*}

\begin{table*}[t]
\setlength{\abovecaptionskip}{0cm}
\setlength{\belowcaptionskip}{0cm}
\caption{Overall performance of FELLRec and other traditional recommendation baselines. Bold signifies the best performance among all methods. Underlined values indicate the second best. * denotes statistically significant improvements of the best method over the second-best, according to the t-tests with a significance level of $p$ \textless 0.01.}
\begin{adjustbox}{width={\linewidth},keepaspectratio}
\label{tab:main_exp_trad}
\begin{tabular}{cl|cccc|cccc|cccc}
\toprule
\multicolumn{2}{c|}{\multirow{2}{*}{\textbf{Method}}}                               & \multicolumn{4}{c|}{\textbf{Games}}                                                   & \multicolumn{4}{c|}{\textbf{Microlens}}                                                        & \multicolumn{4}{c}{\textbf{Book}}                                                         \\
\multicolumn{1}{l}{}                                           & \multicolumn{1}{l|}{} & \textbf{R@10}   & \textbf{R@20}   & \textbf{N@10}     & \textbf{N@20}      & \textbf{R@10}   & \textbf{R@20}   & \textbf{N@10}     & \textbf{N@20}      & \textbf{R@10}   & \textbf{R@20}   & \textbf{N@10}     & \textbf{N@20}     \\ \midrule
\multicolumn{1}{l|}{\multirow{2}{*}{\textbf{Centralized}}} & \textbf{MF}                & 0.0101               & 0.0164               & 0.0070                & 0.0090           & 0.0044               & 0.0063               & 0.0026               & 0.0032                 & 0.0050                & 0.0089               & 0.0060                & 0.0071               \\
\multicolumn{1}{l|}{}                                          & \textbf{LightGCN}                     & 0.0153               & 0.0234               & 0.0101               & 0.0127       & 0.0078               & 0.0116               & 0.0044               & 0.0055               & 0.0065               & \uline{0.0120}                & 0.0078               & 0.0093               \\ \midrule
\multicolumn{1}{l|}{\multirow{5}{*}{\textbf{Federated}}}   & \textbf{FedMF}& 0.0065               & 0.0108               & 0.0044               & 0.0058     & 0.0029               & 0.0045               & 0.0021               & 0.0027                & 0.0050                & 0.0070                & 0.0034               & 0.0041               \\
\multicolumn{1}{l|}{}                                          & \textbf{LightFR}                & 0.0088               & 0.0139               & 0.0051               & 0.0069         & 0.0041               & 0.0055               & 0.0024               & 0.0044          & 0.0048               & 0.0079               & 0.0049               & 0.0061               \\
\multicolumn{1}{l|}{}                                          & \textbf{FedPerGNN}                & 0.0145               & 0.0229               & 0.0093               & 0.0121            & 0.0043               & 0.0060                & 0.0024               & 0.0029        & 0.0062               & 0.0112               & 0.0075               & 0.0089               \\
\multicolumn{1}{l|}{} & \textbf{BIGRec+Ours}             & \uline{0.0158}      & \uline{0.0274}      & \uline{0.0104}      & \uline{0.0139}     & \uline{0.0088}      & \uline{0.0128}      & \uline{0.0051}      & \uline{0.0062}    & \uline{0.0085}      & 0.0102      & \uline{0.0124}      & \uline{0.0116}      \\
\multicolumn{1}{l|}{} & \textbf{RecFormer+Ours}           & \textbf{0.0215*}      & \textbf{0.0373*}      & \textbf{0.0122*}      & \textbf{0.0170*}    & \textbf{0.0141*}      & \textbf{0.0245*}      & \textbf{0.0065*}      & \textbf{0.0094*}       & \textbf{0.0274*}      & \textbf{0.0411*}      & \textbf{0.0275*}      & \textbf{0.0301*} \\ \bottomrule
\end{tabular}
\end{adjustbox}
\end{table*}

\subsection{Experimental Settings}\label{sec:setting}


\subsubsection{Datasets and Settings}
We assess the effectiveness of FELLRec on three popular benchmark datasets. 
1) 
\textbf{Games}\footnote{\url{https://nijianmo.github.io/amazon/index.html.}} is from the Amazon review datasets, which covers interactions between users and video games with rich textual features such as game titles and categories.
2) \textbf{MicroLens}\footnote{\url{https://github.com/westlake-repl/MicroLens.}} is a newly released short video recommendation dataset. Each short video contains raw modal information such as title, cover image, audio, and video information. 
3) \textbf{Book} is also derived from Amazon review datasets, containing users’ interactions with extensive books, encompassing a broad spectrum of genres and subjects.
The datasets' statistics are detailed in Table~\ref{tab:datasets}.
For all three datasets, we organize user-item interactions chronologically based on timestamps and divide the data into training, validation, and testing sets in an 8:1:1 ratio. 

Within the context of LLM-based recommendations, we explore two distinct fine-tuning approaches:
1) \textbf{Few-shot fine-tuning} fine-tunes LLMs using a limited number of examples, \eg 1024-shot. 
2) \textbf{Full fine-tuning} utilizes all samples to fine-tune LLMs.

For the evaluation, we adopt full-ranking protocol~\cite{he2020lightgcn} and evaluate using Recall@$K$ and NDCG@$K$, where $K=10$ or $20$. 

\subsubsection{\textbf{Baselines}}
We compare FELLRec against competitive baselines. First,
we select two superior backend LLMs: 
1) \textbf{BIGRec}~\cite{bao2023bi}.
2) \textbf{RecFormer}~\cite{li2023text}.
Given the absence of LLM-based privacy-preserving recommendation method in existing literature, we incorporate two well-established federated learning algorithms that can be deployed on LLM:
3) \textbf{FedAvg}~\cite{mcmahan2017communication}.
4) \textbf{FedProx}~\cite{li2020federated}.
5) \textbf{Ditto}~\cite{li2021ditto}.
6) \textbf{RoLoRA}~\cite{chen2024robust}.
Additionally, our comparison also includes baselines from traditional recommendation methods. 
Specifically, we select \textbf{MF}~\cite{2011wbpr} and \textbf{LightGCN}~\cite{he2020lightgcn} as the centralized-based method, along with their federated counterparts: \textbf{FedMF}~\cite{chai2020secure}, \textbf{LightFR}~\cite{zhang2023lightfr}, and \textbf{FedPerGNN}~\cite{wu2022federated}.
Details of baselines and implementation are shown in Appendix~\ref{appendix:baselines} and~\ref{appendix:hyper-param_setting}.


\subsection{Overall Performance }\label{sec:overall}
We compare FELLRec with other baselines, shown in Table~\ref{tab:main_exp} and Table~\ref{tab:main_exp_trad}. The result indicates that:
1) FELLRec outperforms other privacy-preserving methods on all datasets and achieves performance on par with centralized LLM-based methods. This efficacy is largely due to the dynamic balance strategy, which offers dynamic parameter aggregation and learning speeds. 
2) FedAvg and FedProx performance fluctuates due to their inability to robustly adapt to varied data distributions across clients and the heterogeneity within clients (see detailed analysis in Appendix~\ref{appendix:in-depth}). Conversely, FELLRec consistently excels, aided by its dynamic balance strategy.
3) FELLRec outperforms all traditional baselines in both centralized and federated settings due to the contextual comprehension and abundant pre-trained knowledge of LLMs, along with the dynamic balance strategy.


\begin{figure}
\setlength{\abovecaptionskip}{-0.1cm}
\setlength{\belowcaptionskip}{0.05cm}
  \centering 
  \hspace{-0.2in}
  \subfigure{
    \includegraphics[width=1.5in]{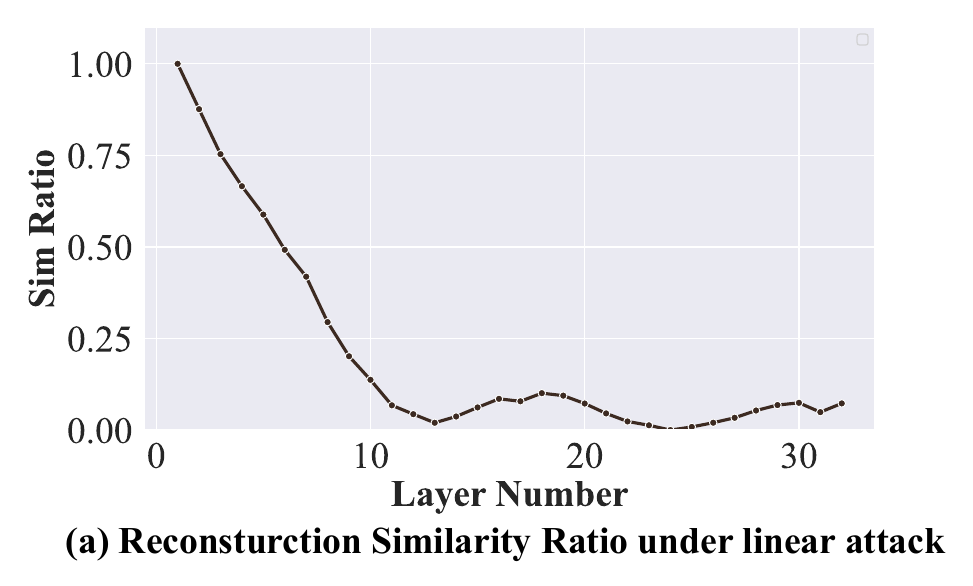}} 
  \hspace{-0.02in}
  \subfigure{
     \includegraphics[width=1.5in]{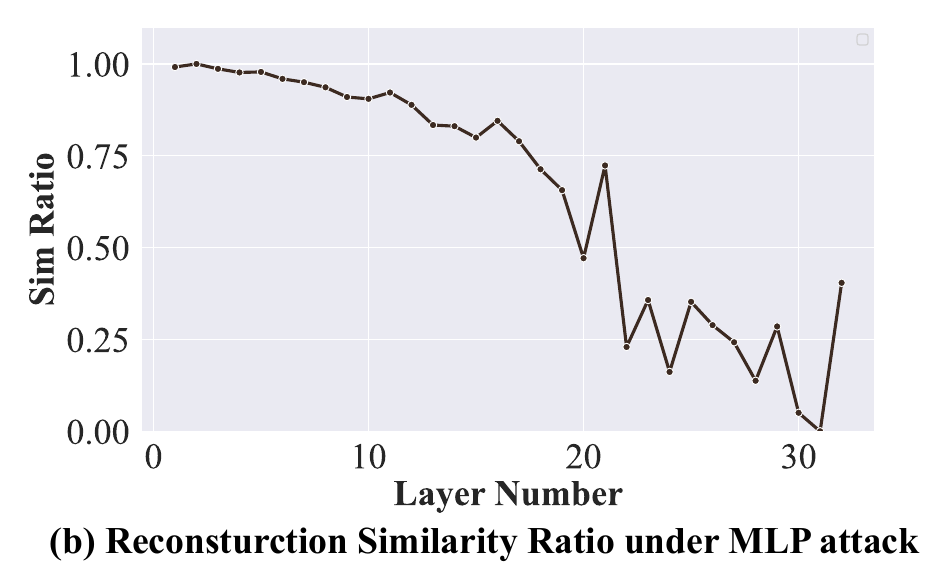}} 
\caption{(a) and (b) shows the similarity between input embeddings and predicted input embeddings according to extracted embeddings of different layers from BIGRec under linear probe attack and MLP probe attack.}
  \label{fig:recover}
\end{figure}

\subsection{Attack Model Analysis}
\label{sec:tradeoff}
To mitigate client-side resource consumption, FELLRec employs the flexible storage strategy in Section~\ref{sec:storage}.
However, putting some layers on the server side may also bring the risk of attacking to leak the privacy.
In this section, we conduct an attack simulation experiment to assess the possibility of attacks via intermediate embeddings processed on the server side.

We use BIGRec as a case study, extracting intermediate output embeddings from all layers and applying two typical types of white-box attack methods: the linear probe attack and the Multilayer Perceptron (MLP) probe attack~\cite{kim2024propile}. These methods attempt to reconstruct the input embedding from the layer embeddings separately~\cite{xu2023llms}.
The detail of the selected attack models is shown in Appendix~\ref{appendix:attack}.
We report the cosine similarity ratio between the reconstructed embeddings and the ground truth input embeddings, as illustrated in Figure~\ref{fig:recover}.
We find that:
1) the likelihood of reconstructing user historical interactions from intermediate embeddings decreases with ascending layer number generally.
2) The possibility of reconstruction from the last layer increases since LLM training aims to align the final output with the target interacted item, which may have higher similarity with the input embeddings.
Thus, the choice of parameter $k$ should be guided by this attack simulation (in this case, $k \geq 21$).
\begin{figure}
\setlength{\abovecaptionskip}{-0.1cm}
\setlength{\belowcaptionskip}{0.05cm}
  \centering 
  \hspace{-0.2in}
  \subfigure{
    \includegraphics[width=1.5in]{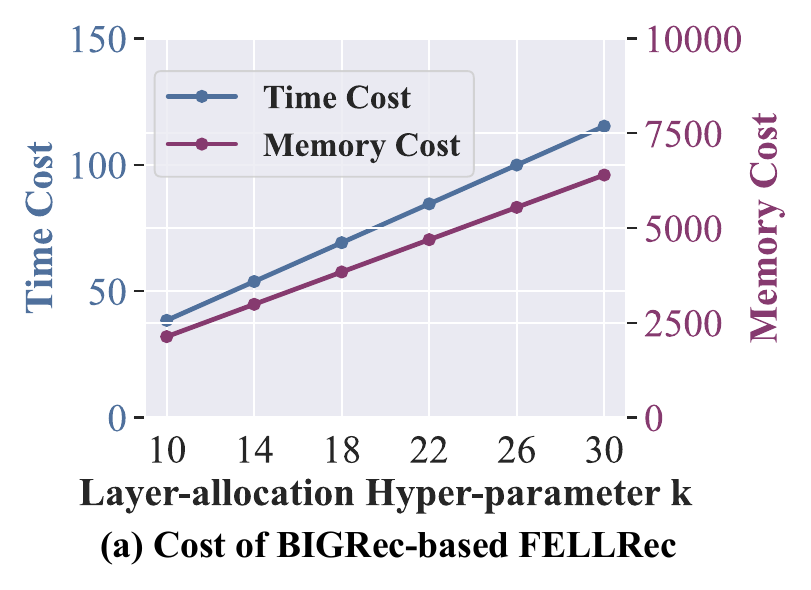}} 
  \hspace{-0.02in}
  \subfigure{
     \includegraphics[width=1.5in]{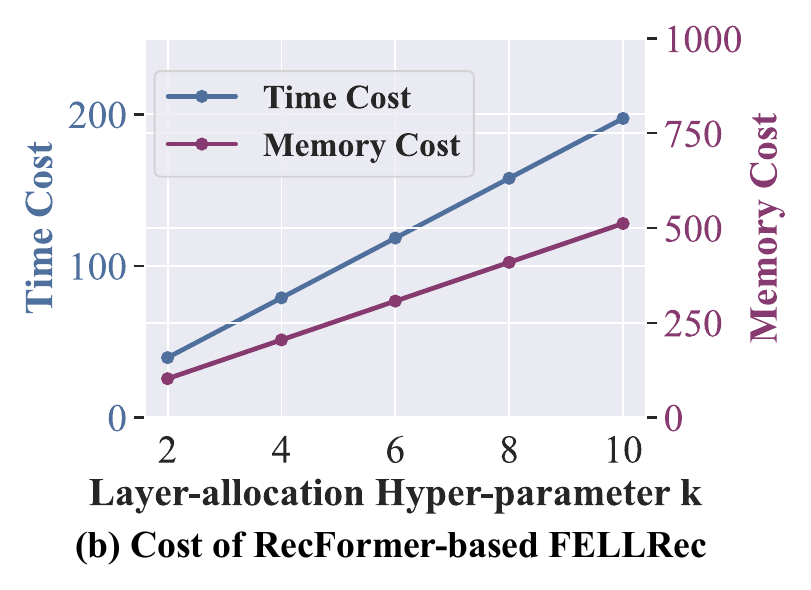}} 
\caption{(a) and (b) show the time (s) and memory (MiB) cost for different values of $k$ under the BIGRec-based and RecFormer-based FELLRec, respectively.}
  \label{fig:efficiency}
\end{figure}

\begin{table}[t!]
\setlength{\abovecaptionskip}{-0.0cm}
\setlength{\belowcaptionskip}{-0.0cm}
\caption{Cost of FELLRec and FedAvg. $k$ and $N$ represent layer-allocation hyper-parameter and LLM layer number, where $k+1<N$. $b$ and $c$ represent the communication cost of uploading one layer of LLM parameters and data embeddings to the server, respectively.}
\begin{adjustbox}{width={\linewidth},keepaspectratio}
\label{tab:cost}
\begin{tabular}{cccc}
\hline
                & \textbf{Storage Cost} & \textbf{Inference Cost} & \textbf{Communication Cost} \\ \hline
\textbf{FELLRec}   & $\mathcal{O}(k+1)$                & $\mathcal{O}(k+1)$                   & $\mathcal{O}((k+1) \cdot b + 2 \cdot c)$                  \\
\textbf{FedAvg} & $\mathcal{O}(N)$                  & $\mathcal{O}(N)$                    & $\mathcal{O}(N \cdot b)$                      \\ \hline
\end{tabular}
\end{adjustbox}
\end{table}

\begin{table*}[t]
\setlength{\abovecaptionskip}{0cm}
\setlength{\belowcaptionskip}{0cm}
\caption{Client evaluation results of the centralized method, FedAvg and FELLRec. Bold represents the lowest degree of imbalance among the methods evaluated, using the same backend model.}
\begin{adjustbox}{width={\linewidth},keepaspectratio}
\label{tab:client_eval}
\begin{tabular}{l|cccccc|cccccc}
\hline
\multirow{2}{*}{\textbf{Recall@10}} & \multicolumn{6}{c|}{\textbf{Games}}                                                                                           & \multicolumn{6}{c}{\textbf{MicroLens}}                                                                                                 \\
                                    & \textbf{Client 1} & \textbf{Client 2} & \textbf{Client 3} & \textbf{Client 4} & \textbf{Client 5} & \textbf{Imbalance} & \textbf{Client 1} & \textbf{Client 2} & \textbf{Client 3}    & \textbf{Client 4}    & \textbf{Client 5}    & \textbf{Imbalance} \\ \hline
\textbf{BIGRec}                     & 0.0227            & 0.0338            & 0.0144            & 0.0163            & 0.0153            & 1.35             & 0.0148            & 0.0275            & 0.0059 & 0.0050 & 0.0031     & 4.50  \\ 
\textbf{$\quad$+FedAvg}                     & 0.0157            & 0.0208            & 0.0235            & 0.0085            & 0.0127            & 1.76             & 0.0010             & 0.0047            & 0.0017               & 0.0001               & 0.0004               & 46.00               \\
\rowcolor[HTML]{EEF5FF} 
\textbf{$\quad$+FELLRec}                       & 0.0171            & 0.0211            & 0.0163            & 0.0136            & 0.0152            & \textbf{0.55}             & 0.0170             & 0.0120             & 0.0066               & 0.0042               & 0.0062               & \textbf{3.04}             \\ \hline
\end{tabular}
\end{adjustbox}
\end{table*}

\subsection{Efficiency Analysis}
\label{sec:efficiecny}
We analyze efficiency of FELLRec both experimentally and theoretically. 
First, we calculate the time and memory cost for different layer-allocation hyper-parameter $k$ in Figure~\ref{fig:efficiency}.
This reveals a trade-off: storing more layers server-side raises the risk of attack but reduces client resource cost. 
Thus, clients can dynamically adjust layer allocation based on capacity.

We also evaluate FELLRec against FedAvg across various metrics, including storage cost, communication cost, and local client inference cost. 
The findings in Table~\ref{tab:cost} demonstrate that our method outperforms FedAvg in storage and inference cost.
For communication cost, our method outperforms FedAvg under the conditions: $(k+1) \cdot b + 2 \cdot c < n \cdot b$,
which simplifies to $c<(n-k-1)b/{2}$.
Intuitively, the lower the value of $k$, the greater the likelihood of achieving superior communication efficiency compared to FedAvg. Similarly, a lower value of $c$, indicating a smaller device scope, further enhances the superiority of our method.
This indicates that our method is particularly effective in clients with limited scope, making it ideally suited for user devices.
Moreover, we can further reduce communication costs for transferring model parameters and data by implementing asynchronous updates~\cite{xu2023asynchronous}.

We also conduct a practical analysis of the communication cost associated with uploading a portion of the LLM's parameters to the server. 
During each communication round, only the LoRA parameters from the client side are uploaded to the server, which significantly reduces the overall parameter size.
Since the size of the LoRA parameters for LLaMA-7B is approximately 16 MB~\cite{hu2021lora}, the communication cost remains manageable. 
To provide clarity, we quantified the communication cost of uploading LoRA parameters. For LLaMA-7B, the full LoRA parameter size is approximately 16 MB per client. 
Assuming a 100 Mbps network (a common configuration), the upload time is approximately 1.28 seconds~\cite{mcmahan2017communication}. 
However, with our Flexible Storage Strategy, only the saved layer parameters from clients need to be uploaded, further reducing the communication cost. 
Moreover, considering that a typical training iteration for LLaMA-7B takes approximately 110 seconds per batch in our experiment, this communication time is negligible in comparison to the overall training time.

\subsection{\textbf{Client Performance Analysis}}
To assess whether FELLRec mitigates the performance imbalance among various clients, we conduct client evaluation experiment, as detailed in Table~\ref{tab:client_eval}.  
Similar results are seen with the Book dataset, but figures are omitted for brevity. 
The imbalance degree is calculated as follows:
$\text{Imbalance Degree} = (m_{\text{best}}-m_{\text{worst}})/m_{\text{worst}},$
where $m_{\text{best}}$ is the Recall@10 of the best client, and $m_{\text{worst}}$ is the Recall@10 of the worst client.
The results indicate that:
1) FELLRec effectively mitigates the performance imbalance issue among clients compared to FedAvg, primarily due to the dynamic balance strategy, which customizes dynamic parameter aggregation and learning speed for different clients.
2) The imbalance degree in the MicroLens dataset is more pronounced under FedAvg, which is potentially caused by the data distribution among clients being much more diverse than others.
Such diversity may lead to conflict optimization objectives among clients, thus exacerbating the imbalance.

\vspace{2pt}
\noindent\textbf{$\bullet$ More In-depth Experiments.}
Ablation study, client heterogeneity analysis, client number study, and hyper-parameter analysis are in Appendix~\ref{appendix:in-depth}.
\section{Related Work}
\label{sec:related_work}
\textbf{$\bullet$ LLM-based Recommendation.}
Recent advances in LLMs for recommendation systems have gained attention for their contextual understanding and pre-trained knowledge~\cite{lin2024data}.
Early efforts, such as P5~\cite{geng2022recommendation}, TALLRec~\cite{bao2023tallrec}, focused on fine-tuning LLMs with prompts and recommendation data. 
Later works, like BIGRec~\cite{bao2023bi} and TIGER~\cite{rajput2023recommender}, refine LLMs by grounding outputs in real item spaces and enhancing generative processes with semantic information.
This shift moves from simply integrating recommendation data to fully leveraging LLMs for improved performance.
As performance improves, the focus expands to include trustworthiness, such as fairness and explainability~\cite{zhang2023chatgpt,wang2023llm4vis}. 
Studies like \cite{xu2023llms, LLM4FairSurvey, Xu-TaxRank-SIGIR24} highlight user unfairness in LLM-based recommendation, and LLMHG~\cite{chu2024llm} introduces an explainable framework combining LLM reasoning with hypergraph neural networks. 

\vspace{2pt}
\noindent\textbf{$\bullet$ Federated Recommendation.}
Fed4Rec enhances data privacy in recommendation systems using federated learning~\cite{yin2024device,zhang2024transfr}, operating under two main frameworks:
1) \textbf{Peer-Peer Framework}~\cite{yang2022dpmf,long2023model,long2023decentralized}: 
Clients directly broadcast intermediate parameters to other clients, who aggregate them into their models.
For example, SemiDFEGL\cite{qu2023semi} improves scalability via device-to-device collaboration, while DGRec~\cite{zheng2023decentralized} uses a decentralized graph neural network. However, this framework has high communication costs due to large LLM parameters.
2) \textbf{Client-Server Framework}~\cite{wang2022fast,liu2023privaterec,imran2023refrs,zhang2023dual}: 
Clients send local parameters to a central server for aggregation and redistribution. 
Examples include FedPerGNN\cite{wu2022federated}, which incorporates high-order information while preserving privacy, and LightFR~\cite{zhang2023lightfr}, a lightweight federated matrix factorization framework with efficient inference.


\section{Conclusion}
\label{sec:conclusion}
In this work, we proposed a federated framework for LLM-based recommendation (FELLRec). 
Firstly, we identified two key challenges in directly applying Fed4Rec in the LLM-based recommendation: exacerbated client performance imbalance and high client resource costs. 
Subsequently, to address these, FELLRec introduces:
1) dynamic balance strategy, which designs dynamic parameter aggregation and learning speed for different clients during training, aims to ensure relatively equitable performance across clients. 
2) Flexible storage strategy, which selectively retains certain sensitive LLM layers on the client side, while offloading other layers to the server, aims to save resources.
Overall, FELLRec offers an equitable and resource-efficient approach to safeguard data privacy in LLM-based recommendations. 

\section*{Acknowledgments}
\label{sec:Acknowledgments}

This research/project is supported by the National Research Foundation Singapore and DSO National Laboratories under the AI Singapore Programme (AISG Award No: AISG2-RP-2020-018)
\clearpage
\section*{Limitations}
First, the largest model we use in this work is LLaMA-7B, exploring the potential of using even larger LLMs could provide further insights into the effectiveness of our method. 
Second, we primarily utilize two common types of white-box attack methods for model analysis. 
Additionally, we only use BIGRec as a case study for these attacks. 
However, different LLMs may demonstrate varying levels of resilience to different attack methods. 
In the future, it would be beneficial to apply a broader range of attack methods across various LLM architectures to further validate the effectiveness of our approach.
Third, while the dynamic balance strategy we designed for FELLRec has proven effective, it would be promising to explore more fine-grained aggregation strategies (\eg layer-based aggregation) in the future.

\bibliography{custom}
\clearpage
\appendix
\section{Appendix}
\label{sec:appendix}




\subsection{Baselines}\label{appendix:baselines}
1) \textbf{BIGRec}~\cite{bao2023bi} using LLaMA-7B as the LLM backbone, utilizing the item title to present the user sequence.
2) \textbf{RecFormer}~\cite{li2023text} using LongFormer as the LLM backbone, utilizing both item titles and descriptions to represent user sequences.
3) \textbf{FedAvg}~\cite{mcmahan2017communication} aggregates client model parameters without uploading their data. 
4) \textbf{FedProx}~\cite{li2020federated} extends FedAvg by adding a proximity term to the local optimization, allowing for more robust handling of heterogeneous data across clients.
5) \textbf{Ditto} is a personalized federated learning framework that simultaneously ensures fairness and robustness in statistically heterogeneous networks via a scalable solver.
6) \textbf{RoLoRA} employs an alternating minimization approach for LoRA to enhance robustness against reduced fine-tuning parameters and heightened data heterogeneity.
7) \textbf{MF}~\cite{2011wbpr} is a classical matrix factorization (MF) approach.
8) \textbf{LightGCN}~\cite{he2020lightgcn} leverages high-order neighbor information to enhance the user and item representations.
9) \textbf{FedMF}~\cite{chai2020secure} is a privacy-enhanced MF approach based on secure homomorphic encryption.
10) \textbf{LightFR}~\cite{zhang2023lightfr} is a lightweight federated recommendation framework with privacy-preserving MF.
11) \textbf{FedPerGNN}~\cite{wu2022federated} designs a privacy-preserving graph expansion protocol to incorporate high-order information under privacy protection in GNN-based recommendation.

\subsection{Implementation}\label{appendix:hyper-param_setting}
For all the baselines, we follow the original settings in their paper for implementation.
Besides, we adopt the parameter-efficient fine-tuning technique LoRA to fine-tune BIGRec in 1024-shot and fully fine-tune RecFormer.
For the client partition, we set the client number equal to 5, and cluster users based on pre-trained MF user embeddings, leveraging the premise that users with analogous characteristics and preferences are more likely to congregate in similar areas or platforms. 
For FedAvg, FedProx and FELLRec, we set the same local round number to ensure a fair comparison. 
The best hyper-parameters are selected with the searching scopes as follows: 
speed-related warm-up factor and time-related warm-up factor are tuned in $\{0.1, 0.3, 0.5, 0.7, 0.9, 1.1, 1.3\}$ and $\{1, 3, 5, 10, 15, 20\}$.
The experiments were conducted under four NVIDIA V100.

\subsection{\textbf{Attack Model Selection}}\label{appendix:attack}
For the attack model, there are two kinds of attack models: white-box~\cite{gurnee2023language} and black-box~\cite{papernot2017practical}. White-box means the attacker has complete knowledge of the target model, including the model's architecture, parameters, and training data. Black-box means attackers have very limited knowledge of the target model. They may only be able to speculate about the model's partial behavior through its inputs and outputs.
To validate the robustness of our approach, we conducted experiments involving two typical types of white-box attacks: the linear probe attack and the Multilayer Perceptron (MLP) probe attack~\cite{kim2024propile}. These methods have better attack capabilities than black-box methods due to their access to more prior knowledge.

The linear probe attack involves training a linear regression model on the intermediate output embeddings of LLM. This model attempts to recover the original data from these intermediate embeddings. We then compare the similarity ratio between the reconstructed embeddings and the ground truth input embeddings, where a lower similarity ratio indicates greater difficulty for the attack model to extract useful information.
The MLP probe attack is similar to the linear probe attack but employs a Multilayer Perceptron (MLP) instead of a simple linear regression model to probe the intermediate representations.

\begin{table}[t!]
\setlength{\abovecaptionskip}{0cm}
\setlength{\belowcaptionskip}{0cm}
\caption{Statistics of three datasets.}
\label{tab:datasets}
\scalebox{0.8}{%
\begin{tabular}{ccccc}
\toprule
\textbf{Dataset}     & \textbf{\#User} & \textbf{\#Item} & \textbf{\#Interaction} & \textbf{Density} \\ \midrule 
\textbf{MicroLens} & 45,886         & 12,413         & 332,730               & 5e-04         \\

\textbf{Games}    & 50,532         & 16,857          & 452,894               & 5e-04         \\

\textbf{Book}  & 64,989         & 56,394         & 4,963,757              & 1.3e-03 \\
\bottomrule
\end{tabular}
}
\end{table}

\subsection{In-depth Experimental analysis}\label{appendix:in-depth}
\begin{figure}
\setlength{\abovecaptionskip}{-0.2cm}
\setlength{\belowcaptionskip}{-0.2cm}
  \centering 
  \hspace{-0.2in}
  \subfigure{
    \includegraphics[width=1.5in]{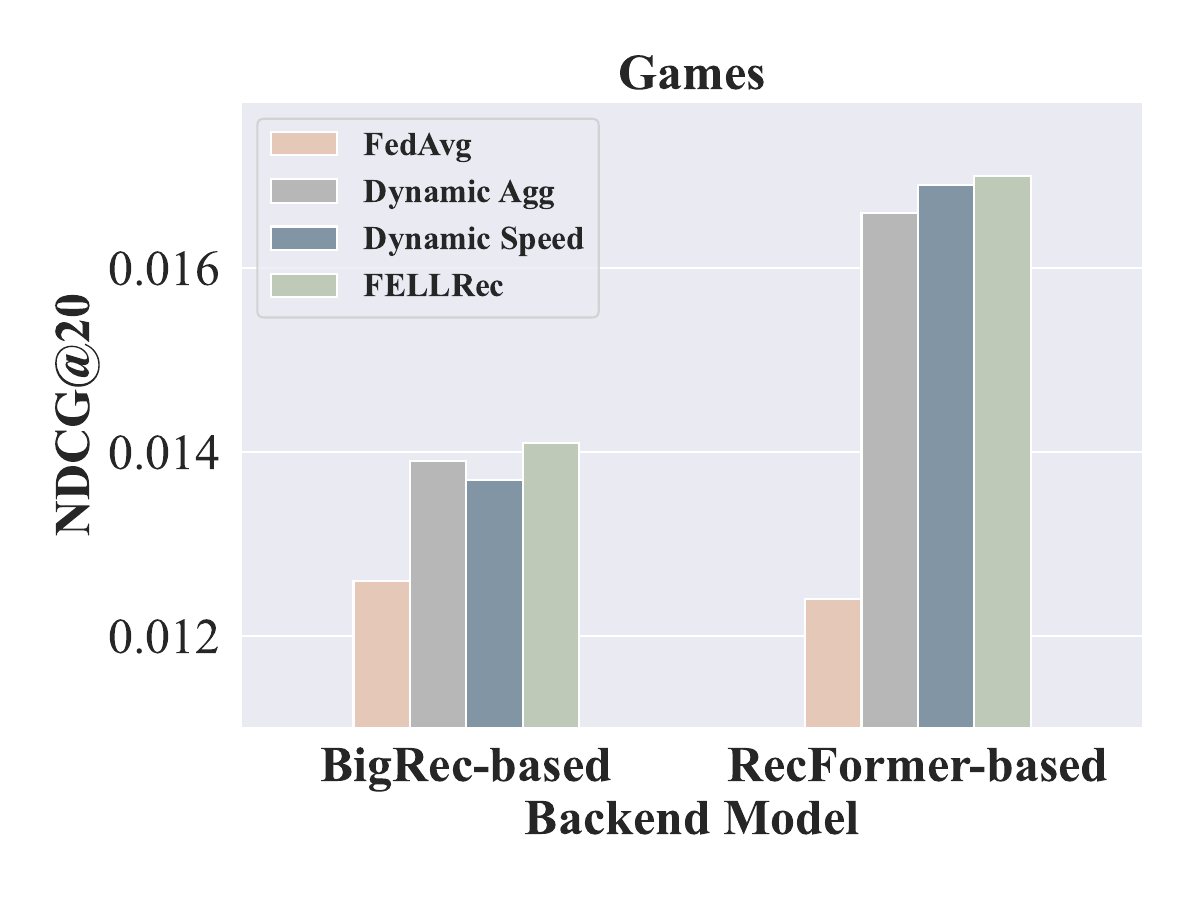}} 
  \hspace{-0.02in}
  \subfigure{
     \includegraphics[width=1.5in]{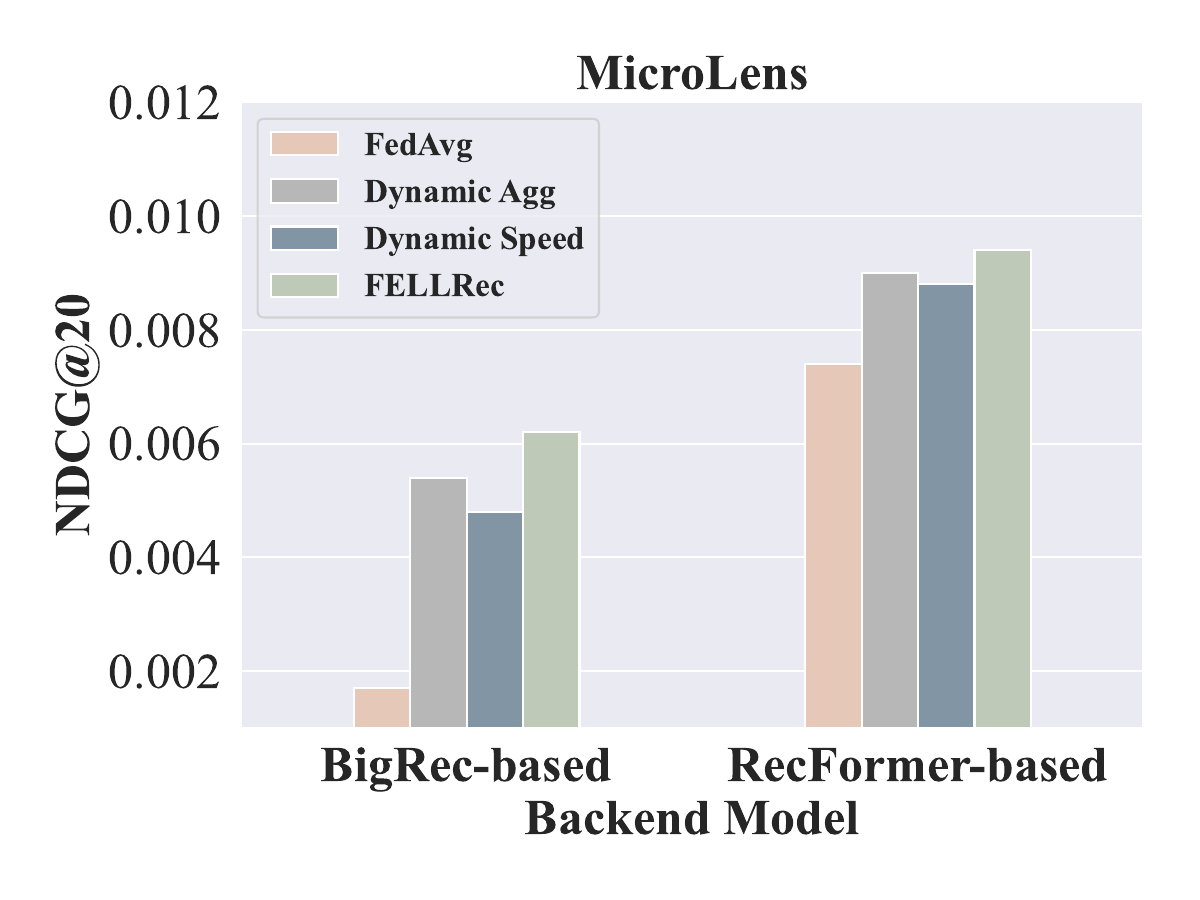}} 
\caption{Contributions of dynamic parameter aggregation and learning speed to FELLRec.}
  \label{fig:abl}
\end{figure}

\subsubsection{\textbf{Ablation Study}}
In this section, we evaluate the unique contributions of dynamic parameter aggregation and dynamic learning speed in comparison with FedAvg, presenting the results in Figure~\ref{fig:abl} for the Games and Microlens datasets (excluding the Book dataset due to analogous trends). 
The analysis reveals that:
1) FELLRec with dynamic parameter aggregation consistently surpasses FedAvg. This highlights the benefits of an attention-based parameter aggregation method that tailors aggregation to the specific data distribution within FELLRec.
2) Similarly, FELLRec with dynamic learning speed invariably outperforms FedAvg, emphasizing the advantages of customizing the learning speed of each client based on their learning status.
3) The effectiveness of the two parts is consistent across different datasets and backend models, further demonstrating their robustness and generalizability.


\begin{figure}
\setlength{\abovecaptionskip}{-0.20cm}
\setlength{\belowcaptionskip}{-0.2cm}
  \centering 
  \hspace{-0.2in}
  \subfigure{
    \includegraphics[width=1.5in]{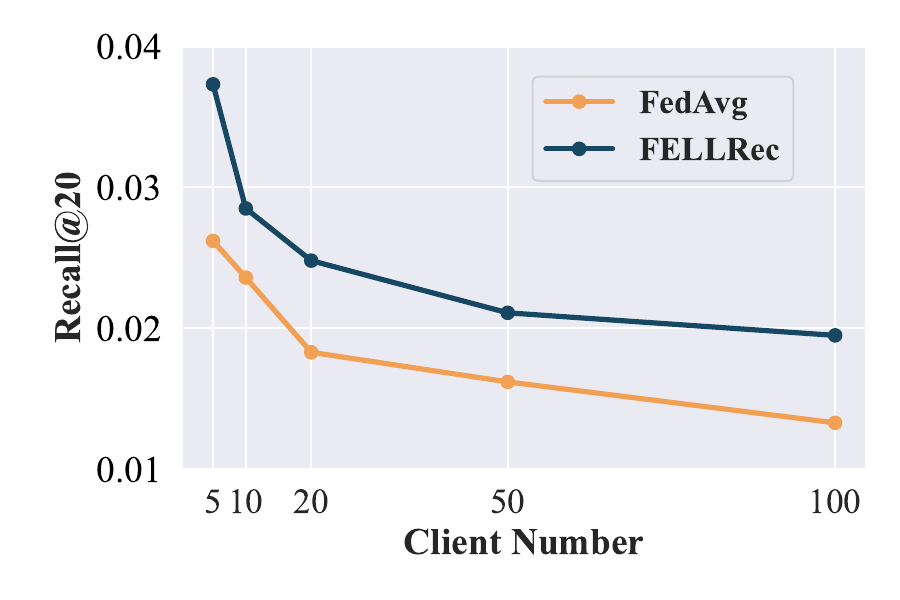}} 
  \hspace{-0.02in}
  \subfigure{
     \includegraphics[width=1.5in]{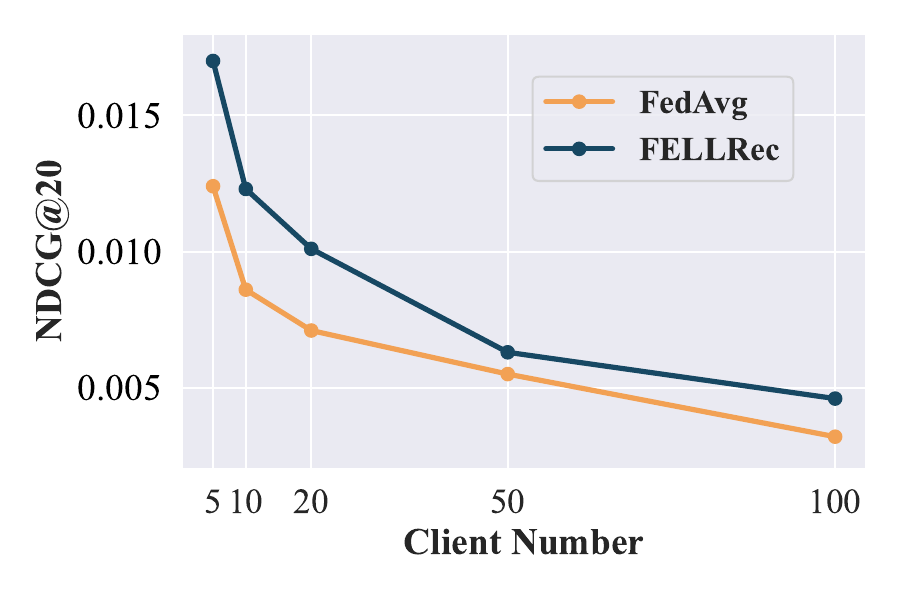}} 
\caption{Influence of Client Number to RecFormer-based FELLRec and FedAvg in Games.}
  \label{fig:number}
\end{figure}

\begin{table}[t!]
\setlength{\abovecaptionskip}{0cm}
\setlength{\belowcaptionskip}{0cm}
\caption{Performance (R@10) of FedAvg and FELLRec under different heterogeneity degrees in Games. As concentration parameter $c$ increases, the heterogeneity degree decreases.}
\label{tab:degree}
\scalebox{0.8}{%
\begin{tabular}{cccccc}
\toprule
        \textbf{$c$} & \textbf{0.1} & \textbf{0.3} & \textbf{0.5} & \textbf{0.7} & \textbf{1} \\ \midrule 
        \textbf{FedAvg} & 0.0134 & 0.0139 & 0.0141 & 0.0157 & 0.0162 \\ 
        \textbf{FELLRec} & 0.0187 & 0.0211 & 0.0212 & 0.0216 & 0.0228 \\
\bottomrule
\end{tabular}
}
\end{table}

\subsubsection{\textbf{Influence of Client Heterogeneity Degree}}
\label{sec:noniid}
To further demonstrate the robustness of our framework under different data distributions, we conduct additional experiments to analyze the influence of the heterogeneous data distribution degree across clients on the performance of federated learning baselines and our proposed method.
 Specifically, we follow the prevailing strategy~\cite{wang2023federated} and distribute samples to all clients based on the Dirichlet distribution.
 We report the performance under different heterogeneity degrees in Table~\ref{tab:degree}, with $c$ as the concentration parameter determining the heterogeneity degree across clients. Intuitively, as the concentration parameter $c$ increases, the heterogeneity degree decreases.
 The results indicate that:
1) FELLRec consistently outperforms FedAvg when using the same backend model (BIGRec), demonstrating the superior capability of the dynamic balance strategy under different heterogeneity degree settings.
2) As the heterogeneity degree increases, the performance of FedAvg decreases, corroborating the conclusion from Section~\ref{sec:overall} that the heterogeneity degree significantly impacts FedAvg’s performance, causing fluctuations under different data distributions across clients.

\subsubsection{\textbf{Influence of Client Number}}
To demonstrate the scalability of our approach with an increased number of clients, we expanded the client count from 5 to 100 and reported the comparative results of FELLRec and FedAvg in Figure~\ref{fig:number}. 
The findings indicate that:
1) With the escalation in client numbers, there is a noticeable decline in the performance of both FELLRec and FedAvg, likely due to the amplified diversity acorss client data distribution, which in turn aggravates the imbalance and adversely affects overall performance.
2) Nevertheless, FELLRec consistently outperforms FedAvg in every client count scenario. This enhanced performance is attributed to the dynamic aggregation strategy employed by FELLRec, effectively countering the imbalances stemming from the varied data distributions among clients.

\begin{figure}
\setlength{\abovecaptionskip}{-0.20cm}
\setlength{\belowcaptionskip}{-0.2cm}
  \centering 
  \hspace{-0.2in}
  \subfigure{
    \includegraphics[width=1.5in]{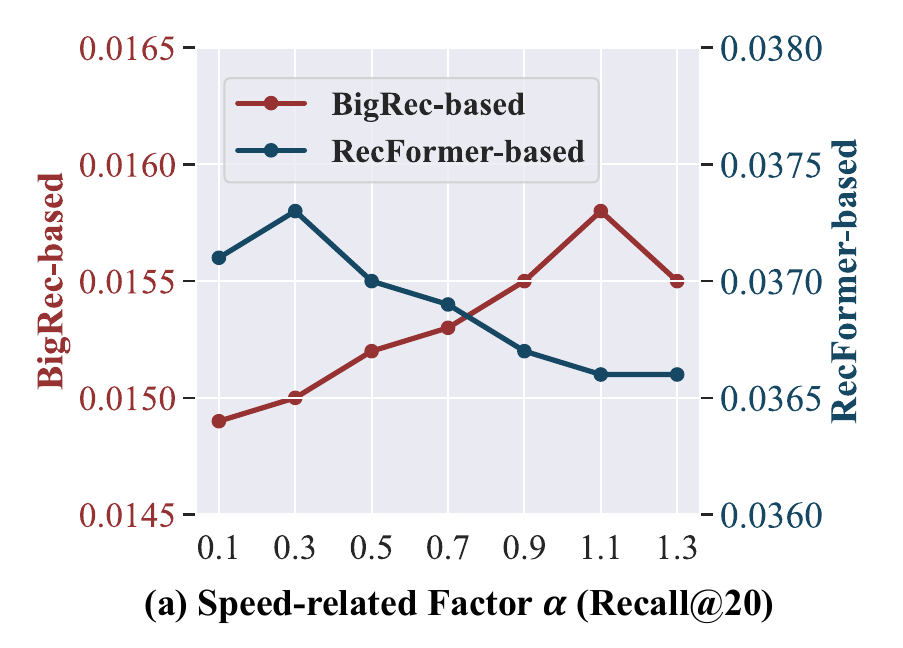}} 
  \hspace{-0.02in}
  \subfigure{
     \includegraphics[width=1.5in]{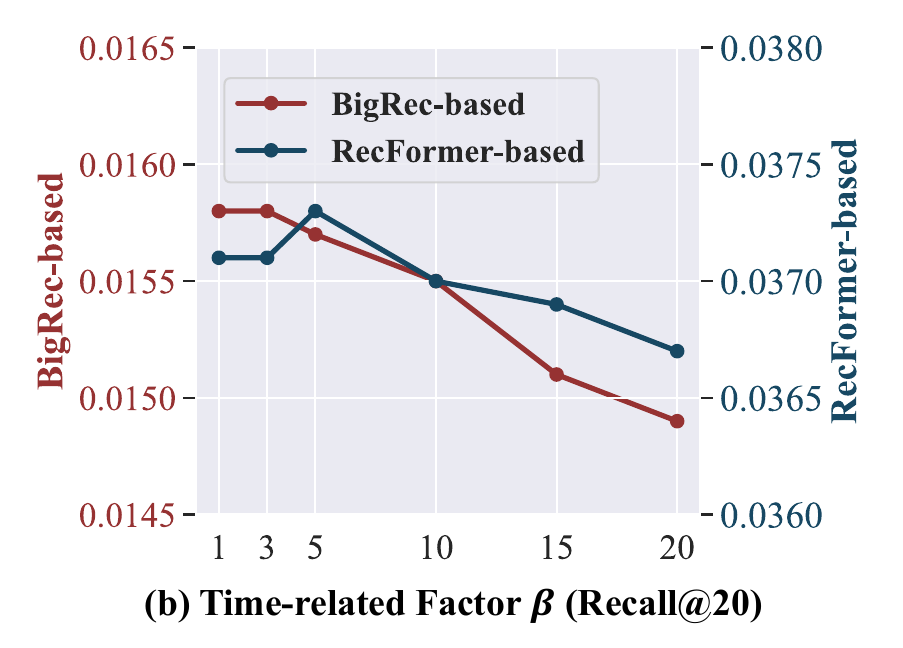}} 
\caption{Hyper-parameters analysis of Games.}
  \label{fig:hyper}
\end{figure}

\subsubsection{\textbf{Hyper-parameter Analysis}}
We select sensitive hyper-parameters, adjusting them within the ranges delineated in Section~\ref{sec:setting}.
The experiment outcomes are visually represented in Figure~\ref{fig:hyper}. 
From our observations:
The settings of the speed-related warm-up factor $\alpha$ and the time-related warm-up factor $\beta$ significantly affect the warm-up speed. Generally, enhancing the values of $\alpha$ and $\beta$ leads to improved performance, facilitating the integration of parameters from other clients to aid the learning process of the current client once it has adequately learned from its data. 
Nevertheless, overly aggressive acceleration in warm-up may prematurely incorporate parameters from other clients before the current client is prepared, potentially disrupting the learning trajectory and adversely affecting performance.

\end{document}